\pdfoutput=1  

\def\acap{\\ \nonumber \\}

\def\rfr#1{Equation\,(\ref{#1})}
\def\rfrs#1#2{Equations\,(\ref{#1})--(\ref{#2})}
\def\Rfr#1{Equation\,(\ref{#1})}
\def\Rfrs#1#2{Equations\,(\ref{#1})--(\ref{#2})}
\def\derp#1#2{\frac{\partial{#1}}{\partial{#2}}}

\def\dert#1#2{\frac{{{\mathit{d}}}{#1}}{{{\mathit{d}}}{#2}}} 
 
\def\eqi{\begin{equation}}
\def\eqf{\end{equation}}
\def\eqia{\begin{eqnarray}}
\def\eqfa{\end{eqnarray}}
\def\rp#1#2{\frac{#1}{#2}}
\def\lb#1{\label{#1}}

\def\nk{n_\mathrm{K}}

\def\ton#1{\left(#1\right)}
\def\qua#1{\left[#1\right]}
\def\grf#1{\left\{#1\right\}}
\def\ang#1{\left\langle #1\right\rangle}

\documentclass[a4paper, 11pt]{article}

\usepackage[english]{babel}
\usepackage[LGR,T1]{fontenc}
\usepackage[utf8]{inputenc}
\usepackage{w-greek}
\usepackage[nointegrals]{wasysym}
\usepackage{authblk}
\usepackage{abstract}
\usepackage{amsmath}
\usepackage{amsthm}
\usepackage{amscd}
\usepackage{amssymb}
\usepackage{dsfont}
\usepackage{xcolor}
\usepackage{graphicx}
\usepackage{epsfig}
\usepackage{xr-hyper}
\usepackage[pdfstartview=XYZ,
bookmarks=true,
colorlinks=true,
linkcolor=blue,
urlcolor=blue,
citecolor=blue,
pdftex,
bookmarks=true,
linktocpage=true, 
hyperindex=true
]{hyperref}
\usepackage{starfont}
\usepackage{txfonts}
\usepackage[mathlines]{lineno}
\usepackage{tabularx}
\usepackage{booktabs}
\usepackage{fontawesome5}

\usepackage[round]{natbib} 
\allowdisplaybreaks[1]

\makeatletter
\DeclareRobustCommand\ref{%
    \@ifstar\@refstar\T@ref
  }%
  \DeclareRobustCommand\pageref{%
    \@ifstar\@pagerefstar\T@pageref
  }%
 \makeatother

\DeclareSymbolFont{greekletters}{LGR}{\familydefault}{m}{n}
\DeclareMathSymbol{\qA}{\mathord}{greekletters}{65}
\DeclareMathSymbol{\qB}{\mathord}{greekletters}{66}
\DeclareMathSymbol{\qG}{\mathord}{greekletters}{71}
\DeclareMathSymbol{\qD}{\mathord}{greekletters}{68}
\DeclareMathSymbol{\qE}{\mathord}{greekletters}{69}
\DeclareMathSymbol{\qZ}{\mathord}{greekletters}{90}
\DeclareMathSymbol{\qEt}{\mathord}{greekletters}{72}
\DeclareMathSymbol{\qTh}{\mathord}{greekletters}{74}
\DeclareMathSymbol{\qI}{\mathord}{greekletters}{73}
\DeclareMathSymbol{\qK}{\mathord}{greekletters}{75}
\DeclareMathSymbol{\qL}{\mathord}{greekletters}{76}
\DeclareMathSymbol{\qM}{\mathord}{greekletters}{77}
\DeclareMathSymbol{\qN}{\mathord}{greekletters}{78}
\DeclareMathSymbol{\qX}{\mathord}{greekletters}{88}
\DeclareMathSymbol{\qO}{\mathord}{greekletters}{79}
\DeclareMathSymbol{\qP}{\mathord}{greekletters}{80}
\DeclareMathSymbol{\qR}{\mathord}{greekletters}{82}
\DeclareMathSymbol{\qS}{\mathord}{greekletters}{83}
\DeclareMathSymbol{\qT}{\mathord}{greekletters}{84}
\DeclareMathSymbol{\qU}{\mathord}{greekletters}{85}
\DeclareMathSymbol{\qPh}{\mathord}{greekletters}{70}
\DeclareMathSymbol{\qCh}{\mathord}{greekletters}{81}
\DeclareMathSymbol{\qPs}{\mathord}{greekletters}{89}
\DeclareMathSymbol{\qOm}{\mathord}{greekletters}{87}
\DeclareMathSymbol{\qa}{\mathord}{greekletters}{97}
\DeclareMathSymbol{\qb}{\mathord}{greekletters}{98}
\DeclareMathSymbol{\qg}{\mathord}{greekletters}{103}
\DeclareMathSymbol{\qd}{\mathord}{greekletters}{100}
\DeclareMathSymbol{\qe}{\mathord}{greekletters}{101}
\DeclareMathSymbol{\qz}{\mathord}{greekletters}{122}
\DeclareMathSymbol{\qet}{\mathord}{greekletters}{104}
\DeclareMathSymbol{\qth}{\mathord}{greekletters}{106}
\DeclareMathSymbol{\qi}{\mathord}{greekletters}{105}
\DeclareMathSymbol{\qk}{\mathord}{greekletters}{107}
\DeclareMathSymbol{\ql}{\mathord}{greekletters}{108}
\DeclareMathSymbol{\qm}{\mathord}{greekletters}{109}
\DeclareMathSymbol{\qn}{\mathord}{greekletters}{110}
\DeclareMathSymbol{\qx}{\mathord}{greekletters}{120}
\DeclareMathSymbol{\qo}{\mathord}{greekletters}{111}
\DeclareMathSymbol{\qp}{\mathord}{greekletters}{112}
\DeclareMathSymbol{\qr}{\mathord}{greekletters}{114}
\DeclareMathSymbol{\fs}{\mathord}{greekletters}{99}       
\DeclareMathSymbol{\qs}{\mathord}{greekletters}{115}       
\DeclareMathSymbol{\qt}{\mathord}{greekletters}{116}
\DeclareMathSymbol{\qu}{\mathord}{greekletters}{117}
\DeclareMathSymbol{\qvf}{\mathord}{greekletters}{102}
\DeclareMathSymbol{\qch}{\mathord}{greekletters}{113}
\DeclareMathSymbol{\qps}{\mathord}{greekletters}{121}
\DeclareMathSymbol{\qom}{\mathord}{greekletters}{119}
\DeclareMathSymbol{\sui}{\mathord}{greekletters}{124}
\DeclareMathSymbol{\dig}{\mathord}{greekletters}{147}
\DeclareMathAccent{\uml}{\mathord}{greekletters}{34} 
\DeclareMathAccent{\umld}{\mathord}{greekletters}{35} 
\DeclareMathAccent{\mcr}{\mathord}{greekletters}{31}
\DeclareMathAccent{\aca}{\mathord}{greekletters}{39} 
\DeclareMathAccent{\ga}{\mathord}{greekletters}{96} 
\DeclareMathAccent{\rb}{\mathord}{greekletters}{60} 
\DeclareMathAccent{\smb}{\mathord}{greekletters}{62} 
\DeclareMathAccent{\ca}{\mathord}{greekletters}{126} 
\DeclareMathAccent{\carb}{\mathord}{greekletters}{64} 
\DeclareMathAccent{\casb}{\mathord}{greekletters}{92} 
\DeclareMathAccent{\aarb}{\mathord}{greekletters}{86} 
\DeclareMathAccent{\aasb}{\mathord}{greekletters}{94} 
\DeclareMathAccent{\garb}{\mathord}{greekletters}{67} 
\DeclareMathAccent{\gasb}{\mathord}{greekletters}{95} 

\DeclareTextSymbol{\textquoteleft}{LGR}{28}      
\DeclareTextSymbol{\textquoteright}{LGR}{29}     

\definecolor{orcidlogocol}{rgb}{0.65, 0.807, 0.223}

\newcommand{\orcid}[1]{$\,$\href{https://orcid.org/#1}{\textcolor{orcidlogocol}{\faOrcid}}}

\title{
\textbf{
\textcolor{black}{Can the jet precession of M87$^\ast$ be caused by a distant intermediate-mass black hole?}
}
}

\author[]{
Lorenzo Iorio\,\orcid{0000-0003-4949-2694}
}

\affil[]{
\href{https://ror.org/01ehyh486}{Ministero dell' Istruzione e del Merito}
\\
Viale Unit\`{a} di Italia 68, I-70125, Bari, Italy \\ email: \href{mailto:lorenzo.iorio@libero.it}{\texttt{lorenzo.iorio@libero.it}}
}

\date{\today}

\providecommand{\keywords}[1]{keywords--- #1}

\begin{document}

\maketitle

\begin{center}
\begin{abstract}
\noindent
The long-term rates of change of all the Keplerian orbital elements of a two-body system acted upon by a remote massive pointlike object are analytically worked out, to the Newtonian quadrupolar level, without any restriction either on the orbital eccentricity and inclination of the disturbed pair or the position of the distant perturber. The latter is considered fixed during the average over one orbital revolution of the inner binary by means of which its orbital perturbations are calculated. The results are presented in a compact form that facilitates a straightforward application \textcolor{black}{to the case of the supermassive black hole-or megapyknon-M87$^\ast$.  In principle, the presence of another distant intermediate-mass black hole may concur to cause the observed jet precession, assumed tightly coupled with the accretion disk. Such a possibility is ruled out by the exclusion plots in the parameter space obtained with the approach presented here}.
\end{abstract}
\end{center}


\section{Introduction}
Let a gravitationally bound two-body system $\mathcal{S}$ composed of a pair of bodies whose total mass and
relative separation are $M$ and $r$, respectively, be considered. If a pointlike object of mass $M^{'}$ slowly revolves about $\mathcal{S}$ at a distance $r^{'}$ larger enough to neglect terms of order higher than the quadrupolar one in the multipolar expansion of its additional potential $\Delta U^{'}_\mathrm{pert}$,
the otherwise elliptical path of the relative motion\footnote{This assumption implies that its unperturbed trajectory is considered as purely Keplerian. Thus, the consequences of the post-Newtonian part of the field of $\mathcal{S}$ and of possible Newtonian deviations from spherical symmetry of it are neglected; differently stated, \textcolor{black}{both bodies} of $\mathcal{S}$ are assumed to be spherical, and general relativity is neglected in their relative motion.} of the components of $\mathcal{S}$ is cumulatively changed because of the gravitational pull of such a distant perturber.

\textcolor{black}{The aim} of the present paper is to analytically calculate such orbital variations, to the Newtonian quadrupole\footnote{For  a calculation of the Newtonian perturbations to higher perturbative orders, see, e.g., \citet{2021PhRvD.103f3003W,2024PhRvD.110h3022C}.} order, without any a-priori restricting assumption on either the orbital configuration of the internal motion of $\mathcal{S}$ or the position of its distant perturber. In particular, the orbital perturbations averaged over one orbital period  of the inner binary, during which $M^{'}$ will be considered essentially fixed in space, will be obtained. 
This scenario becomes relevant in all those cases where the data extend over a time interval that covers a significant number of orbital revolutions of $\mathcal{S}$, while remaining significantly shorter than the orbital period of the third body. Furthermore, even if the results obtained here refer to the Newtonian order, they may well be crucial also if post-Newtonian effects are looked for. Indeed, in all those cases where general relativistic $N$-body features of motion are the aim of the research \citep{2025PhRvD.112h4020F}, the classical perturbations treated here represent a major source of systematic bias in their hoped measurement.

It is certainly not the first time that the problem of the  disturbances of a two-body system due to a tertiary is addressed in the literature, even in the limit of large distances of the latter from the inner binary; see, e.g., \citet{1962AJ.....67..591K,1962P&SS....9..719L,2000ssd..book.....M,2016ARA&A..54..441N}, and references therein. Nonetheless, the resulting formulas generally have a limited validity since they are tailored just to the particular system at hand or are written in a way which is not always easily  understandable by readers not acquainted with the particular problem for which they were purposely derived. In general, people addressing these scenarios are interested in features like their long-term stability and possible chaotic character. Last but not least, often only the effects on some orbital elements, of interest for the specific context considered, have been calculated; they are often the eccentricity and, sometimes, the inclination as well. On the contrary, the expressions obtained here, covering all the six Keplerian orbital elements, can be  applied to a variety of contexts of interest in current astronomical and astrophysical research where the scenario under consideration frequently occurs. Furthermore, they can be easily specialized to specific orbital configurations of both the perturber and the perturbed bodies. 

In particular, the following case will be examined in order to effectively show the wide extent of the results of the present work. The hypothesis that a distant intermediate-mass black hole
\citep{2019MNRAS.488L..90S} may be responsible of the recently observed precession of the jet emanating from the neighbourhood of the supermassive black hole-or megapyknon; see \citet{2025Univ...11..251I}-M87$^\ast$ \citep{2023Natur.621..711C,2025NatAs...9.1218C}. 

The paper is organized as follows. The calculational approach is reviewed in Section \ref{sec:2}. The long-term rates of change of the binary's relative orbit due to a distant perturbing tertiary are analytically worked out in Section \ref{sec:3}. Section \ref{sec:5} deals with the orbital precession of the disk-jet around M87$^\ast$. Section\,\ref{sec:7} summarizes the findings and offers conclusions.
\section{The calculational scheme}\lb{sec:2}
For the sake of clarity, the binary's orbit will be parameterized in terms of the osculating Keplerian orbital elements \citep{1961mcm..book.....B,1991ercm.book.....B,Kaula00,2000ssd..book.....M,2003ASSL..293.....B,2005som..book.....C,2005ormo.book.....R,2008orbi.book.....X,2011rcms.book.....K,2016ASSL..436.....G,SoffelHan19}. They are as follows. The semimajor axis $a$ determines the size of the unperturbed ellipse, the eccentricity $e$ characterizes the shape of the latter, the inclination $I$ measures the tilt of the orbital plane to the fundamental one assumed as reference $\grf{x,y}$ plane, the longitude of the ascending node $\mathit{\Omega}$ is an angle reckoned in the fundamental plane from the reference $x$ direction to the line\footnote{It is the intersection of the orbital plane with the fundamental plane.} of nodes towards the ascending node $\ascnode$, the argument of pericentre $\omega$ is an angle in the orbital plane counted from $\ascnode$ to the pericentre's position, and the mean anomaly at epoch $\eta$ is proportional to the time of passage at pericentre $t_\mathrm{p}$. 

The unperturbed orbital plane of $\mathcal{S}$ is spanned by the two mutually orthogonal unit vectors \citep{Sof89,1991ercm.book.....B,SoffelHan19}
\begin{align}
\boldsymbol{\hat{l}} \lb{elle}& = \grf{\cos\mathit{\Omega},\sin\mathit{\Omega},0}, \acap
\boldsymbol{\hat{m}} \lb{emme}& = \grf{-\cos I\sin\mathit{\Omega},\cos I\cos\mathit{\Omega},\sin I}.
\end{align}
While $\boldsymbol{\hat{l}}$ is directed towards $\ascnode$ along the line of nodes, $\boldsymbol{\hat{m}}$ lies in the orbital plane itself perpendicularly to $\boldsymbol{\hat{l}}$  so that
\eqi
\boldsymbol{\hat{l}}\boldsymbol\times\boldsymbol{\hat{m}}=\boldsymbol{\hat{h}}.\lb{crosso}
\eqf
In \rfr{crosso}, the unit vector $\boldsymbol{\hat{h}}$, defined as
\eqi
\boldsymbol{\hat{h}} \lb{acca} = \grf{\sin I\sin\mathit{\Omega},-\sin I\cos\mathit{\Omega},\cos I},
\eqf
is directed along the binary's orbital angular momentum perpendicularly to the orbital plane.

When the effects of some gravitational dynamical features 
can be expressed in terms of a perturbing potential $\Delta U_\mathrm{pert}\ton{\boldsymbol r}$ depending only on the position  vector $\boldsymbol r$ of the binary's relative orbit, the rates of change of its Keplerian orbital elements, averaged over one orbital revolution, can be  calculated with the planetary equations  in the form of Lagrange. They are \citep{1961mcm..book.....B,1991ercm.book.....B,Kaula00,2000ssd..book.....M,2003ASSL..293.....B,2005som..book.....C,2005ormo.book.....R,2008orbi.book.....X,2011rcms.book.....K,2016ASSL..436.....G,SoffelHan19}
\begin{align}
\ang{\dert a t} \lb{Lag_a} & = \rp{2}{\nk a}\derp{\ang{\mathfrak{R}}}{\eta}, \acap
\ang{\dert e t} \lb{Lag_e} & = \rp{1 - e^2}{\nk a^2 e}\derp{\ang{\mathfrak{R}}}{\eta} - \rp{\sqrt{1 - e^2}}{\nk a^2 e}\derp{\ang{\mathfrak{R}}}{\omega},\acap
\ang{\dert I t} \lb{Lag_I} & = \rp{\cot I}{\nk a^2\sqrt{1 - e^2}}\derp{\ang{\mathfrak{R}}}{\omega} - \rp{\csc I}{\nk a^2\sqrt{1 - e^2}}\derp{\ang{\mathfrak{R}}}{{\mathit{\Omega}}},\acap
\ang{\dert{\mathit{\Omega}} t} \lb{Lag_O} & = \rp{\csc I}{\nk a^2\sqrt{1 - e^2}}\derp{\ang{\mathfrak{R}}}{I},\acap
\ang{\dert\omega t} \lb{Lag_o} & = -\rp{\cot I}{\nk a^2\sqrt{1 - e^2}}\derp{\ang{\mathfrak{R}}}{I} + \rp{\sqrt{1 - e^2}}{\nk a^2 e}\derp{\ang{\mathfrak{R}}}{e},\acap
\ang{\dert\eta t} \lb{Lag_eta} & = -\rp{1 - e^2}{\nk a^2 e}\derp{\ang{\mathfrak{R}}}{e} - \rp{2}{\nk a}\derp{\ang{\mathfrak{R}}}{a},
\end{align}
where the angular brackets $\ang{\cdots}$ denotes the average over one orbital period of the binary's relative motion, $\nk:=\sqrt{\upmu/a^3}$ is the Keplerian mean motion of its relative orbit, $\upmu:=GM$ is its standard gravitational parameter, being $G$ the Newtonian constant of gravitation, and $\mathfrak{R}$ is the so-called disturbing function corresponding to $\Delta U_\mathrm{pert}$ with a minus sign.
\section{The averaged rates of change of the orbital parameters due to a distant pointlike third body}\lb{sec:3}
The perturbing potential due to a pointlike third body, assumed sufficiently far from the perturbed binary, can be expressed, to the quadrupolar order, as \citep{1991AJ....101.2274H}
\eqi
\Delta U^{'}_\mathrm{pert}\ton{\boldsymbol r} \simeq \rp{\upmu^{'}r^2}{2{r^{'}}^3}\qua{1 - 3\ton{{\boldsymbol{\hat{r}}}^{'}\boldsymbol\cdot\boldsymbol{\hat{r}}}^2}\lb{Hogg}
\eqf
In \rfr{Hogg},  $\upmu^{'}:=GM^{'}$ and ${\boldsymbol{\hat{r}}}^{'}$ are the standard gravitational parameter and the unit position vector of the perturber, respectively. In view of what will follow, it is convenient to express the versor of the position vector of the relative orbit of $\mathcal{S}$ as
\eqi
\boldsymbol{\hat{r}} = \boldsymbol{\hat{l}}\cos\ton{\omega + f} + \boldsymbol{\hat{m}}\sin\ton{\omega +  f}\lb{erre},
\eqf
where $f$ is the time-dependent true anomaly, an angle counted in the orbital plane from the pericentre to the instantaneous position on the unperturbed ellipse.
By defining
\begin{align}
\mathtt{Rl} \lb{ql}& := {\boldsymbol{\hat{r}}}^{'}\boldsymbol\cdot\boldsymbol{\hat{l}}, \acap
\mathtt{Rm} \lb{qm}& := {\boldsymbol{\hat{r}}}^{'}\boldsymbol\cdot\boldsymbol{\hat{m}}, \acap
\mathtt{Rh} \lb{qh}& := {\boldsymbol{\hat{r}}}^{'}\boldsymbol\cdot\boldsymbol{\hat{h}},
\end{align}
the disturbing function corresponding to \rfr{Hogg}, averaged over one orbital period of the disturbed binary during which it is assumed that the far perturber does not appreciably change its position, turns out to be
\eqi
\ang{\mathfrak{R}} = \rp{\upmu^{'} a^2}{8{r^{'}}^3}\grf{\ton{2 + 3 e^2}\qua{-2 + 3\ton{\mathtt{Rl}^2 + \mathtt{Rm}^2}} +
 15 e^2 \qua{\ton{\mathtt{Rl}^2 - \mathtt{Rm}^2}\cos 2\omega + 2 \mathtt{Rl}\,\mathtt{Rm} \sin 2\omega}}.\lb{Uave}
\eqf

Then, the net rates of change of the Keplerian orbital elements of the perturbed binary, calculated by inserting \rfr{Uave} in \rfrs{Lag_a}{Lag_eta}, are
\begin{align}
\ang{\dert a t} \lb{dadt} & = 0, \acap
\ang{\dert e t} \lb{dedt} & = \rp{15 e\upmu^{'}\sqrt{1 - e^2}}{4{r^{'}}^3 n_\mathrm{K}}\qua{-2 \mathtt{Rl}\,\mathtt{Rm} \cos 2\omega + \ton{\mathtt{Rl}^2 - \mathtt{Rm}^2} \sin 2\omega}, \acap
\ang{\dert I t} \lb{dIdt} & = \rp{3\upmu^{'}\mathtt{Rh}}{4{r^{'}}^3 n_\mathrm{K}\sqrt{1 - e^2}}\qua{\ton{2 + 3 e^2} \mathtt{Rl} + 5 e^2 \ton{\mathtt{Rl}\cos 2\omega + \mathtt{Rm}\sin 2\omega}}, \acap
\ang{\dert{\mathit{\Omega}} t} \lb{dOdt} & = \rp{3\upmu^{'}\csc I \mathtt{Rh}}{4{r^{'}}^3 n_\mathrm{K}\sqrt{1 - e^2}}\grf{\ton{2 + 3 e^2} \mathtt{Rm} +  5 e^2\ton{-\mathtt{Rm}\cos 2\omega + \mathtt{Rl}\sin 2\omega}}, \acap
\ang{\dert \omega t} \nonumber & = -\rp{3\upmu^{'}}{4{r^{'}}^3 n_\mathrm{K}\sqrt{1 - e^2}}\grf{\ton{-1 + e^2} \qua{-2 + 3\ton{\mathtt{Rl}^2 + \mathtt{Rm}^2}} + \ton{2 + 3 e^2} \mathtt{Rh}\,\mathtt{Rm} \cot I  \right.\acap
\lb{dodt}&\left. -5\qua{-\ton{-1 + e^2}\ton{\mathtt{Rl}^2 - \mathtt{Rm}^2} + e^2 \mathtt{Rh}\,\mathtt{Rm} \cot I} \cos 2\omega + 5 \mathtt{Rl} \qua{2 \mathtt{Rm} \ton{-1 + e^2} + e^2 \mathtt{Rh} \cot I} \sin 2\omega}, \acap
\ang{\dert \eta t} \nonumber & = -\rp{\upmu^{'}}{4{r^{'}}^3 n_\mathrm{K}}\grf{\ton{7 + 3 e^2} \qua{-2 + 3\ton{\mathtt{Rl}^2 + \mathtt{Rm}^2}} + 15 \ton{1 + e^2}\ton{\mathtt{Rl}^2 - \mathtt{Rm}^2}\cos 2\omega + \right.\acap
\lb{detadt} & \left. + 30 \ton{1 + e^2} \mathtt{Rl}\,\mathtt{Rm}\sin 2\omega}.
\end{align}
In general, \rfrs{dadt}{detadt}
describe genuine secular trends. Indeed, all the orbital elements do vary, apart from $a$,  because \rfrs{dadt}{detadt} are mutually coupled. Furthermore, in any realistic scenario, they can vary also because of other dynamical effects like the primary's oblateness, general relativity, the tug by other major bodies in the system at hand, \textcolor{black}{tides and possibly drag as well}. Thus, slow harmonic modulations generally characterize the temporal patterns of the orbital elements of a binary perturbed by a distant pointlike third body.

In the limit $e\rightarrow 0$, the eccentricity remains constant since \rfr{dedt} vanishes. Instead, the orbital plane is generally shifted; indeed, \rfrs{dIdt}{dOdt} reduce to
\begin{align}
\ang{\dert I t} \lb{dIdt0} & = \rp{3\upmu^{'}\mathtt{Rh}\,\mathtt{Rl}}{2{r^{'}}^3 n_\mathrm{K}}, \acap
\ang{\dert{\mathit{\Omega}} t} \lb{dOdt0} & = \rp{3\upmu^{'}\csc I \mathtt{Rh}\,\mathtt{Rm}}{2{r^{'}}^3 n_\mathrm{K}}.
\end{align}
Formally, \rfrs{dodt}{detadt} do not exhibit any singularity, but $\omega$ and $\eta$ do not retain their meaning in a circular orbit.
%
%
\section{The jet precession in the supermassive black hole M87$^\ast$}\lb{sec:5}
Recently, the precession of the jet emanating from the surrounding of the supermassive black hole (SMBH) at the center of the giant elliptical galaxy M87 was measured with the Very Long Baseline Interferometry (VLBI) technique \citep{2023Natur.621..711C,2025NatAs...9.1218C}. By assuming that the SMBH's accretion disk is tightly coupled with the jet \citep{2013Sci...339...49M,2018MNRAS.474L..81L,2025NatAs...9.1218C}, it was recently shown \citep{2025MNRAS.537.1470I} that most of the observed features of the jet's precessional pattern of M87$^\ast$ can be successfully understood in terms of the first-order post-Newtonian (1pN) precessions of the orbital angular momentum of a fictitious test particle moving along a circular orbit with an effective radius of about 14 gravitational radii\footnote{The gravitational radius is defined as $R_\mathrm{g}:=\upmu/c^2$, where $c$ is the speed of light.} induced by the gravitomagnetic Lense-Thirring and the quadrupolar fields of the hole as predicted by the no-hair theorems\footnote{According to them, the multipolar structure \citep{1970Natur.226...64B} of the field of a rotating Kerr black hole \citep{1963PhRvL..11..237K,2015CQGra..32l4006T} is determined uniquely by its mass and angular momentum.} \citep{1967PhRv..164.1776I,1971PhRvL..26..331C,1975PhRvL..34..905R}. Such a result supports much more complex simulations previously performed in the framework of general relativistic magnetohydrodynamics (GRMHD) by taking into account only the gravitomagnetic field \citep{2023Natur.621..711C}.

A natural alternative or, at least, a contributing cause for the observed jet/disk precession may be a possible companion of M87$^\ast$.
It is more likely that it is a distant intermediate-mass black hole (IMBH) \citep{2019MNRAS.488L..90S} whose mass lies in the range $M^{'} \simeq 10^2-10^5 M_\odot$.
Furthermore, in order not to tidally disrupt the accretion disk around M87$^\ast$, it should orbit at a distance $r^{'}$ from its host given by \citep{2019MNRAS.488L..90S}
\eqi
r^{'}>r_\mathrm{t}:=r_0\ton{1 + \xi^{1/3}},
\eqf
where $r_\mathrm{t}$ is the distance cut-off to avoid the tidal disruption of the accretion disk, $r_0$ is the radius of the latter and
\eqi
\xi:=\rp{M^{'}}{M}
\eqf
is the ratio of $M^{'}$ to the mass $M$ of M87$^\ast$. On the other hand, observations indicate that the center of M87 is highly stable, effectively ruling out the presence even of a moderately massive secondary BH \citep{2009Sci...325..444A}.

\Rfrs{dIdt0}{dOdt0}, valid for $e=0$, can be expressed in compact for as
\eqi
\dert{\boldsymbol{\hat{h}}}t = {\boldsymbol{\Omega}}^{'}\boldsymbol{\times}\boldsymbol{\hat{h}}, \lb{hprec}
\eqf
where the precession velocity vector is defined as
\eqi
{\boldsymbol{\Omega}}^{'} := \rp{3\upmu^{'}}{2{r^{'}}^3\nk}\ton{{\boldsymbol{\hat{r}}}^\mathrm{'}\boldsymbol\cdot\boldsymbol{\hat{h}}}{\boldsymbol{\hat{r}}}^{'}.\lb{Omd}
\eqf
\Rfr{hprec} implies that the orbital angular momentum of the effective test particle undergoes a precession about the (slowly varying) position vector of the distant perturber  with the precession velocity ${\boldsymbol{\Omega}}^{'}$ given by \rfr{Omd}. It should be noted that, whether it lies in the orbital plane or perpendicularly to it, there is no precession of the particle's orbital angular momentum, in agreement with \rfrs{dIdt0}{dOdt0}.

In order to test the hypothesis that a distant IMBH is fully responsible of the observed jet precession in M87$^\ast$, the condition that the magnitude of \rfr{Omd}, viewed as a function of the three variables
\begin{align}
u \lb{ics}&:= \log_{10}\ton{\rp{M^{'}}{M_\odot}}\acap
v & := r^{'} - r_\mathrm{t},\acap
w\lb{zeta}&:= \varphi =\arccos\ton{{\boldsymbol{\hat{r}}}^\mathrm{'}\boldsymbol\cdot\boldsymbol{\hat{h}}},
\end{align}
lies within the experimental range \citep{2023Natur.621..711C}
\eqi
\left|\Omega_\mathrm{exp}\right| =0.56\pm 0.02,\,\mathrm{rad\,yr}^{-1}
\eqf
of the measured absolute value of the jet precession is imposed. More specifically, the constraint
\eqi
0.54\,\mathrm{rad\,yr}^{-1}\leq\left|{\boldsymbol{\Omega}}^{'}\ton{u,v,w}\right|\leq 0.58\,\mathrm{rad\,yr}^{-1}\lb{condiz}
\eqf
allows to exclude all those values of the variables of \rfrs{ics}{zeta} for which \rfr{condiz} is not fulfilled.

\begin{figure}
\centering
\begin{tabular}{c}
\includegraphics[width = 10 cm]{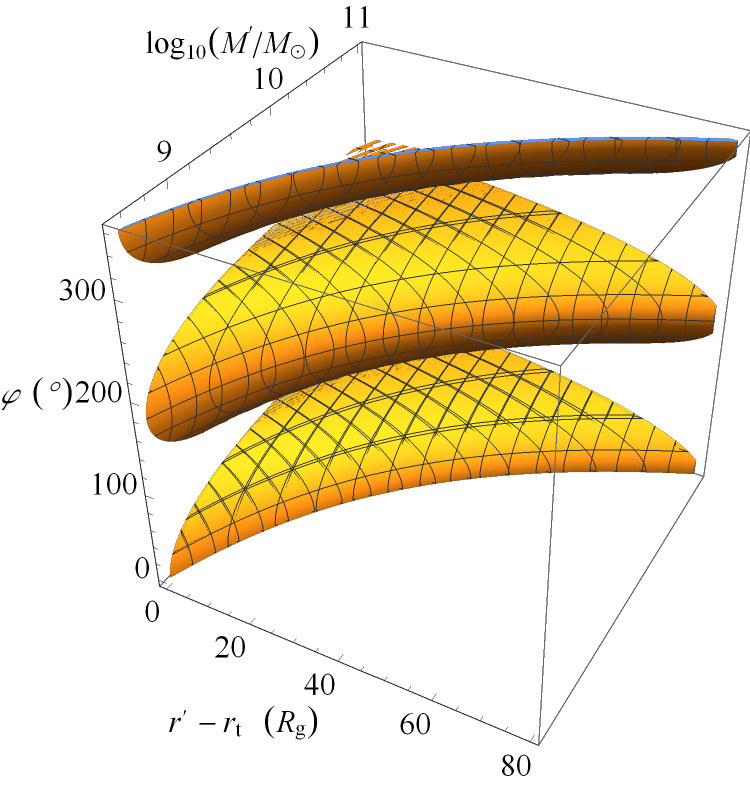}\\
\end{tabular}
\caption{Allowed domains in the 3D parameter space spanned by $u:=\log_{10}\ton{M^{'}/M_\odot}$, $v:=r^{'}-r_\mathrm{t}$, in units of $R_\mathrm{g}$, $w:=\varphi$, in degrees, according to the condition of \rfr{condiz}. The value $r_0=14.1\,R_\mathrm{g}$ is assumed for the effective disk radius.
}\label{fig:prima}
\end{figure}

It turns out that there are no allowed domains in the three-dimensional parameter space spanned by  \rfrs{ics}{zeta}
for
\begin{align}
u &\in\qua{2,5},\acap
v \lb{cond2}&>0,\acap
w \lb{cond3}&\in\qua{0^\circ,360^\circ}.
\end{align}
Indeed, without prejudice to the constraints of \rfrs{cond2}{cond3}, the condition of \rfr{condiz} is fulfilled only for
\eqi
u\in[8.5,11],
\eqf
as shown by Figure \ref{fig:prima} obtained for $r_0=14.1 R_\mathrm{g}$ \citep{2025MNRAS.537.1470I}.
Thus, the hypothesis that an IMBH is the cause of the observed jet precession in M87$^\ast$ can be reasonably ruled out, further strengthening the no-hair theorems.

In this respect, it is important to remark that \rfrs{Lag_I}{Lag_O}, containing $\partial\ang{\mathfrak{R}}/\partial I, \partial\ang{\mathfrak{R}}/\partial\Omega,\partial\ang{\mathfrak{R}}/\partial\omega$,  allow to straightforwardly exclude most of the several long-range exotic models of gravity put forth in recent years (see, e.g., \citet{2007IJGMM..04..115N,2010LRR....13....3D,2010LRR....13....5M,2011PhR...509..167C,2012PhR...513....1C,2012AIPC.1471..103F,2015Univ....1..199C,2016RPPh...79j6901C}, and references therein) as possible causes of the jet precession of M87$^\ast$. Indeed, they generally imply  spherically symmetric extra-potentials $\Delta U_\mathrm{exotic}\ton{r}$.
Since for a Keplerian reference orbit it is
\eqi
r=\rp{a\ton{1-e^2}}{1 + e\cos f},
\eqf
the resulting averaged disturbing function $\ang{\mathfrak{R}_\mathrm{exotic}}$ cannot depend on any of the three orbital elements $I,\mathit{\Omega},\omega$, thus implying $\ang{dI/dt} = \ang{d\mathit{\Omega}/dt} = 0$.
\section{Summary and conclusions}\lb{sec:7}
\textcolor{black}{L}ong-term classical perturbations of all the Keplerian orbital elements of the relative orbit of a gravitationally bound two-body system induced by a distant pointlike perturber were analytically calculated, to the Newtonian quadrupole order, for arbitrary values of the eccentricity and the inclination of the disturbed binary  and for any location of the third body. Averaged over one orbital revolution of the inner pair, during which the third body is assumed essentially fixed in space, the binary's orbital rates of change were expressed in a compact form suitable to be applied to some astronomical scenarios of recent interest. They hold for any orbital configurations of both the perturbed and the perturbing bodies. 

A useful application of the results of this paper was the measured precession of the jet emanating from the billion solar masses megapyknon M$87^\ast$. Despite it was successfully explained in terms of the general relativistic Lense-Thirring precession of the orbital plane-assumed tightly coupled with the jet-of an effective test particle moving in a circular path about 14 gravitational radii wide, an alternative viable candidate would be, in principle, a distant black hole ranging from a hundred to a hundred thousand solar masses. Using the formulas for the inclination and node precessions of a test particle in circular motion derived in this paper allowed \textcolor{black}{to rule out} such a hypothesis, thus providing further support to the general relativistic explanation of the observed phenomenology. \textcolor{black}{The vast majority of all the alternative models of gravity put forth in recent years is ruled out as possible cause of the jet precession of M87$^\ast$ because they leave the orientation of the orbital plane in space unaffected in view of their spherical symmetry.}

The range of applicability of the analytical expressions obtained in the present paper is not limited just to the aforementioned scenario. Indeed, they can be applied also to quite different systems like, e.g., natural or artificial satellites of a planet disturbed by other major bodies, or stars orbiting a megapyknon perturbed by other distant stars as in the Galactic Center. In all those cases where post-Newtonian features of motion are the target, the competing classical perturbations treated here act as a major source of systematic uncertainty, thus requiring a careful modeling. \textcolor{black}{To this aim, it should be remarked that, in principle, such a straightforward comparison may be misleading since when post-Newtonian effects are present, the standard Lagrange planetary equations render results in contact elements, not
in osculating ones. Indeed, under such perturbations, some of the osculating elements
do not coincide with their contact counterparts.}
\section*{Data availability}
No new data were generated or analysed in support of this research.
\section*{Conflict of interest statement}
I declare no conflicts of interest.
\section*{Funding}
This research received no external funding.
\section*{Acknowledgements}
\textcolor{black}{I am grateful M. Efroimsky for important remarks and critical observations.}

\bibliography{Megabib}{}
\end{document}